\documentclass[amsmath,amssymb,reprint,bibnotes]{revtex4-2}
\usepackage{graphicx}
\usepackage{bm}
\usepackage{amsmath} 
\usepackage{graphics}
\usepackage[dvipsnames]{xcolor}
\usepackage[mathlines]{lineno}
\usepackage{braket}
\usepackage[compact]{titlesec}
\begin{document}
\title{All-Optical Control of Ultrafast Plasmon Resonances in the Pulse-Driven Extraordinary Optical Transmission}
\author{Hira Asif$^{\bf (1)}$}
\author{Mehmet Emre Tasgin$^{\bf (2)}$}
\author{Ramazan Sahin$^{\bf (1)}$}\email{rsahin@itu.edu.tr}

\affiliation{${\bf (1)}$ {Department of Physics, Akdeniz University, 07058 Antalya, Turkey}}
\affiliation{${\bf (2)}$ {Institute of Nuclear Sciences, Hacettepe University, 06800 Ankara, Turkey}}

\date{\today}

\begin{abstract}
Understanding the ultrafast processes at their natural-time scale is crucial for controlling and manipulating nanoscale optoelectronic devices under light-matter interaction. Here, we demonstrate that ultrafast plasmon resonances, attributed to the phenomenon of Extraordinary Optical Transmission (EOT), can be significantly modified by tuning the spectral and temporal properties of the ultrashort light pulse. In this scheme, all-optical active tuning governs spatial and temporal enhancement of plasmon oscillations in the EOT system without device customization. We analyze the spectral and temporal evolution of the system through two approaches. First, we develop a theoretical framework based on the coupled harmonic oscillator model, which analytically describes the dynamics of plasmon modes in the coupled and uncoupled state. Later, we compare the evolution of the system under continuous wave and pulsed illumination. Further, we discuss time-resolved spectral and spatial dynamics of plasmon modes through 3D-FDTD simulation method and wavelet transform. Our results show that optical tuning of oscillation time, intensity, and spectral properties of propagating and localized plasmon modes yields a 3-fold enhancement in the EOT signal. The active tuning of the EOT sensor through ultrashort light pulses pave the way for the development of on-chip photonic devices employing high-resolution imaging and sensing of abundant atomic and molecular systems.
\end{abstract}

\maketitle

\section{Introduction} 
All-optical control of plasmon-based nanophotonic systems is highly desirable for on-chip device manipulation, coherent control of photonic signals, ultrafast optical switching, quantum sensing, and high-resolution optical imaging \cite{Dhama2022, Taghinejad2019, Lechago2019}. It also enables active tuning of nanoscale intense electromagnetic fields supported by surface plasmon resonances (SPR) in the metallic nanostructures. SPR has opened new avenues in the realm of optical sensing and nanoscale imaging by introducing the phenomenon of Extraordinary Optical Transmission (EOT) \cite{Zhang2015, Liu2004, Lertvachirapaiboon2018}. The enhanced light transmission from metals designed with periodic nanohole arrays is attributed to the excitation of surface plasmon polaritons (SPP) at the metal-dielectric interface by the incident light. These SPPs in the form of evanescent waves, tunnel through the subwavelength holes, interfere with the SPP at the other interface, and then scatter to the far field \cite{Garcia-Vidal2010, Ebbesen1998}. Apart from bound SPP modes, surface plasmons also excite at the rim of nanohole structure in the form of localized surface plasmons (LSP) \cite{Degiron2005}. Consequently, these optical oscillations (SPP and LSP) form the basis of enhanced transmission, and their dynamical response at the top-bottom interfaces of the film plays a crucial role in the optical properties of zero-order extraordinary transmission of light. Therefore, modulating the spectral, spatial, and temporal properties of plasmonic fields can enable active tuning of EOT signals and make it a promising tool to build highly integrated plasmon-based active photonic devices \cite{Lee2015}.\\ 
In this regard, the main challenge is to control the intrinsic properties of SPP and LSP modes through active elements which can modulate transmission properties without device customization. Fundamentally, the spectral features of SPP and LSP modes depend on the geometry and dimensions of the metal-hole nanostructure \cite{Chen2015, Sangiao2016}. More specifically, the passive elements such as hole diameter, periodicity, and the dielectric environment alter the spectral response of plasmon resonances \cite{Gehan2011}. The spatiotemporal dynamics of plasmon modes including nanoscale intense near-field \cite{Barnes2006} and ultrafast damped oscillations \cite{Barnes2003} can be controlled through external stimuli such as quantum objects and optical light source. For instance, placing a quantum emitter inside the nanohole structure and its coupling with LSP resonances has demonstrated the spectral and spatial modulation of transmitted light and enables low dissipation of the signal during the evolution of the system \cite{Yildiz2020, Sahin2020}.\\
On the other hand, active tuning through an optical light pulse enables coherent control of plasmonic fields owing to their femtosecond coherence timescale. For instance, the control over spatial dispersion of localized plasmon modes has been achieved by using phase-modulated femtosecond light pulse to restrain delocalization \cite{Stockman2004}. Also, the near-field dynamics of the SPP field in the metallic gratings have been studied in the presence and absence of substrate after illumination with femtosecond light source \cite{Roland2018}. These studies have suggested, to probe the time-dependent femtosecond dynamics of plasmon modes in real-time, i.e., femtoseconds, the coherent control of the system requires ultrashort pulse excitation and for the efficient resonant coupling the excitation time should be shorter than the dephasing time of plasmon resonances \cite{Stockman2002}.\\
\begin{figure}[hbt!]
\includegraphics[scale=0.43]{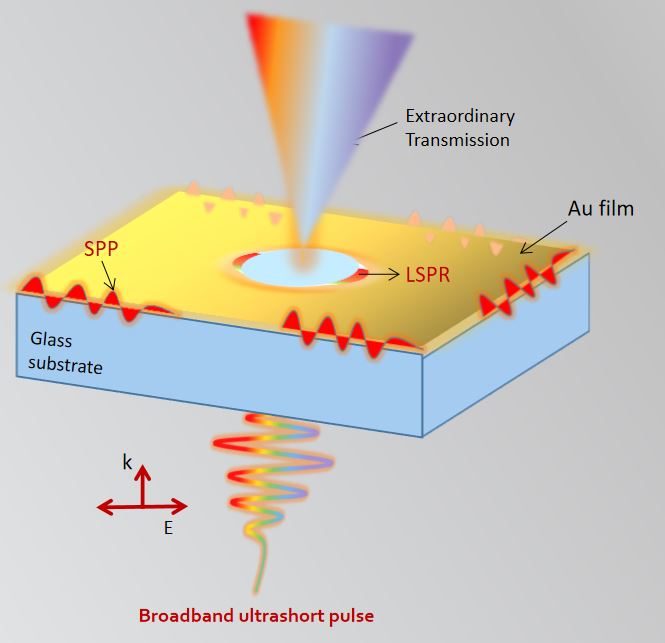}
\caption{\label{Fig1} Schematic representation of EOT device. The device consists of a circular hole of diameter 200 nm in a 100 nm thick Au film laying on a glass substrate with a thin adhesion layer of Cr. The excited SPP and LSP mode are shown at the surface and around the hole driven by ultrashort pulse.}
\end{figure} 
Despite achieving the coherent control over spectral and spatial dynamics, the ultrashort lifetime (10 to 20 fs) of plasmon oscillations \cite{qelife2013} inhibits their potential use in many applications which is still in debate. This ultrafast damping of plasmon polaritons is due to the radiative and non-radiative decay pathways \cite{Giannini2011,Melikyan2004}. In the EOT structure, SPP damping mainly occurs due to the scattering at nanoholes, which makes the hole diameter possible mechanism for radiative damping \cite{Roland2003}. To reduce damping by modifying the geometry of the EOT sensor is practical and challenging. Nevertheless, with fixed geometrical parameters, the optical properties of SPP and LSP can be modified or elevated through all-optical tuning governed by ultrashort light pulses which brings new possibilities in ultrafast active plasmonics \cite{MacDonald2009}.\\
Ultrashort pulses with broad spectral and temporal functional degrees of freedom offer simultaneous excitation of coherent pathways and their interference give rise to transient modulation in plasmonic behavior \cite{Bahar2022}. In comparison, the possibility of active tuning of ultrafast plasmon dynamics in the steady-state through continuous wave (CW) excitation is quite low because of the rapid redistribution of deposited energy through the system in a short time scale \cite{Baumert2007}. Besides most of the EOT studies have been reported under the interaction of CW light with the metal-nanohole array. However, tuning of plasmonic response in EOT structure through femtosecond light pulse and the modulation of spatio-temporal properties simultaneously in its real-time has not been proposed yet.\\ 
In this study, we demonstrate all-optical modulation of SPP and LSP modes driven by ultrashort light pulse in the EOT device (Fig.\ref{Fig1}). In the evolution of the EOT system, the temporal characteristics (pulse shape, duration, and spectral bandwidth) of the driving pulse are used as the functional degrees of freedom for the time-resolved tuning of SPP and LSP resonances. We show that, contrary to conventional CW excitation, temporal tailoring and spectral tuning of ultrashort pulse leads to a significant enhancement in the intense nanoscale field and average lifetime of plasmon modes which strongly modifies the EOT signal exceeding to $95\%$ of transmission from EOT device. We optimize a theoretical framework that evaluates the response of the plasmonic fields for plane wave (CW) and ultrashort pulse excitation independently. Our theory suggests that the enhanced characteristics of plasmon resonances stem from the constructive interference of multiple phase-locked modes driven by tailored light pulse. We further evaluate the average lifetime of plasmon modes by varying the pulse duration and find a direct connection between the optimal pulse durations and extreme ultrafast plasmon damping. To justify our theoretical results, we perform 3D FDTD simulations and explore the broadband pulse driven time-dependent spectral and temporal dynamics of plasmons modes contributing to enhanced transmission. We also investigate the instantaneous frequencies of hybrid plasmon modes in the time-domain through wavelet transform. Such optical control clearly demonstrate the time-resolved manipulation and clear understanding of enhanced transmission, which has been relaying so far on the geometrical manipulation of the EOT device.\\      
\section{Theoretical Model}
Here, we utilize a model system of two coupled harmonic oscillators to investigate the time-dependent spatial and temporal dynamics of plasmon modes and estimate the optical response of the system acting as a plasmon-induced EOT device. For simplicity in the analytical calculations, the EOT system is modeled as a single holey metal film laying on a glass substrate, and the quantized plasmon modes propagating along the surface as SPP and localized at the hole edge as LSP are excited by a TM-polarized broadband light source incident perpendicular to the metal plane in the z-direction (Fig.\ref{Fig1}).
\begin{figure}[hbt!]
\includegraphics[scale=0.31]{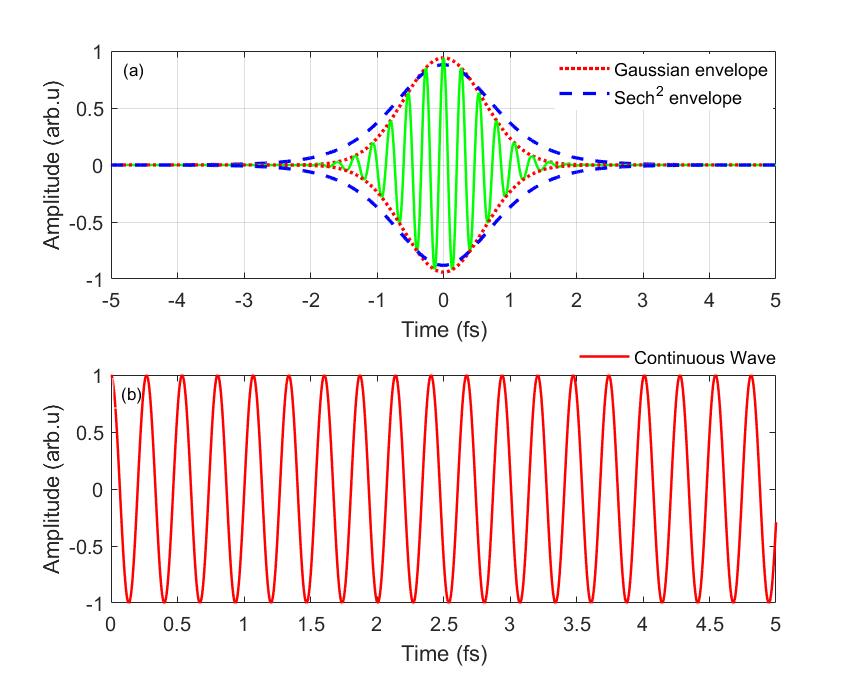}
\caption{\label{Fig2}Temporal waveforms of driving light sources. a) Broadband ultrashort pulse with Gaussian and $Sech^2$ shaped envelopes for $E_o=1$, $\tau=10 fs$, carrier frequency $\omega_{o}=2.51*10^{15} Hz$ with zero carrier phase and no chirp. b) Continuous wave (CW) with the same carrier frequency.}
\end{figure}
We study the nanohole-metal system, for two different excitation sources; at first we illuminate the structure with CW light and examine the steady-state response of both plasmon modes in the coupled and uncoupled state. Later we investigate the pulse response of the system, excited by an ultrashort pulse (shaped with Gaussian or $Sech^2$ envelope functions), in the time-domain. Analytically, a CW light is expressed as a steady function $Ae^{-i\omega_o t}$ with a fixed carrier amplitude A and frequency $\omega_o$ in time 't'. Whereas a good approximation for an ultrashort pulse is the Gaussian pulse given by the analytical function of the form,
\begin{equation}
E(t) = E_o e^{-i\omega_{o}t}*A(t)
\label{eq:1} 
\end{equation}
Here, A(t) is the pulse envelope which is defined as $e^{-2ln2\frac{t^2}{\tau^2}}$ for Gaussian and $sech^2({t/\tau})$ for $Sech^2$ envelope with $\tau$ as pulse duration. The temporal waveforms of CW and femtosecond light pulse are shown in Fig.\ref{Fig2}. In the time-dependent interaction of light with quantum mechanical system, the coherent dynamics of SPP and LSP modes in the coupled plasmonic system are described by the total Hamiltonian $(\hat{\mathcal{H}}_{tot})$ of the system in the Schrodinger picture given as, 
\begin{eqnarray}
\hat{\mathcal{H}}_{tot}=\hbar\omega_{p}\hat{a}_{p}^\dagger\hat{a}_{p}+\hbar\omega_{l}\hat{a}_{l}^\dagger\hat{a}_{l}+\hbar f_{p,l}(\hat{a}_{p}^\dagger\hat{a}_{l}+\hat{a}_{l}^\dagger\hat{a}_{p})\nonumber\\
+i\hbar E_oA(t)e^{-i\omega_o t}(\hat{a}_{p,l}^\dagger)
\label{eq:2}
\end{eqnarray}
where $\hbar\omega_{p}$ and $\hbar\omega_{l}$ are the resonant energies of propagating ($p$) and localized ($l$) normal modes along with, $\hat{a}_{p,l}^\dagger(\hat{a}_{p,l})$, creation(annihilation) operators respectively. The third term demonstrate the coupling between both SPP and LSP with interaction parameter $f_{p,l}$. The forth term in Eq.\ref{eq:2} corresponds to the energy transferred by the free propagating light source to the system. The driven-dissipative dynamics of both quantized modes are derived by solving Heisenberg equations. 
\begin{equation}
\partial_t\hat{a}_{p}= i[\hat{\mathcal{H}},\hat{a}_{p}], \hspace{0.5cm}  \partial_t\hat{a}_{l}= i[\hat{\mathcal{H}},\hat{a}_{l}].
\label{eq:3}
\end{equation} 
After solving Eq.\ref{eq:3}, the equations of motion are derived representing the complex amplitude of propagating ($\alpha_{p}$) and localized plasmon modes ($\alpha_{l}$). 
\begin{equation}
 \dot{\alpha}_{p}= -(i\omega_{p}+\gamma_{p})\alpha_{p}+E_o e^{-i\omega_o t} A(t)-ig\alpha_{l}
 \label{eq:4}
\end{equation}
\begin{equation}
 \dot{\alpha}_{l}= -(i\omega_{l}+\gamma_{l})\alpha_{l}+E_o e^{-i\omega_o t} A(t)-ig\alpha_{p}
 \label{eq:5}
\end{equation}
\begin{figure}[hbt!]
\includegraphics[scale=0.25]{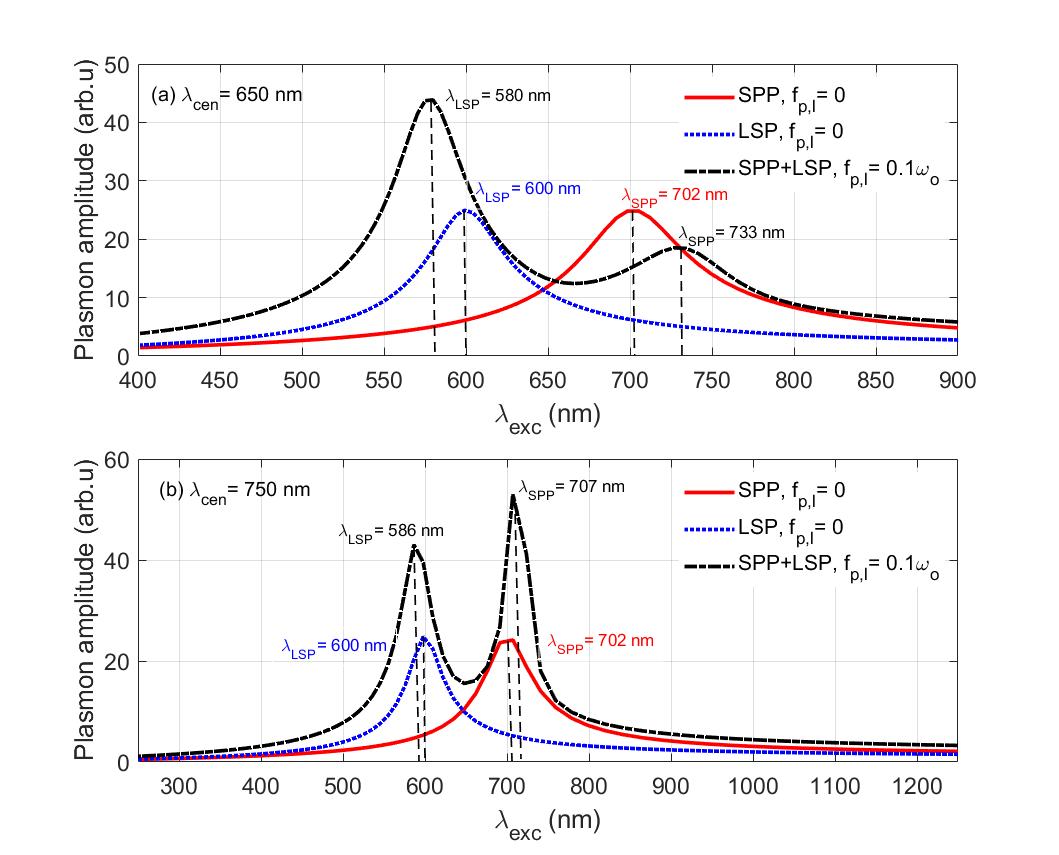}
\caption{\label{Fig3} Steady-state amplitudes of plasmon modes as a function of excitation wavelength for two different driving sources, (a) carrier wavelength $650$ nm, $\omega_{p}= 0.93\omega_o$, $\omega_{l}= 1.08\omega_o$, and (b) $750$ nm, $\omega_{p}= 1.07\omega_o$, $\omega_{l}= 1.25\omega_o$ and $\gamma_{p,l}=0.04\omega_o$ with $f_{p,l}=0.1\omega_o$ and without coupling $f_{p,l}=0$.}
\end{figure} 
where $\gamma_{p}$ and $\gamma_l$ are the damping rates of SPP and LSP modes, respectively. Typically, the plasmon oscillations have large decay rate $\approx 10^{14}Hz$ \cite{guenay_controlling_2020}. We calculate the resonant wavelength of SPP mode, i.e., 702 nm, through the dispersion relation \cite{Barnes2006}, for a structure consisting of 400 nm periodic array of 200 nm-sized circular hole in a 100 nm thick Au film layered on a glass substrate. For an isolated hole of diameter 200 nm, the resonance wavelength of LSP mode is taken as 600 nm \cite{Nehl2008}. In the EOT structure, the interaction between SPP and LSP modes is very strong due to the nanoscale distances separating the periodic indentation. Such interaction leads to the hybridization of plasmon modes near the hole edge due to ultrafast energy and polarization exchange. In the strong coupling regime, the interaction parameter $f_{p,l}$ is taken as $0.1\omega_o$. To investigate the spectral response of plasmon modes, we choose two different carrier wavelengths $(\lambda_{cen})$, 650 nm and 750 nm, of broadband light source. For simplicity in the theoretical calculations, all the parameters of the system, i.e. decay rates and the frequencies of plasmon modes (mentioned in Fig.\ref{Fig3}) are taken in the units of excitation frequency $\omega_o$ corresponding to $(\lambda_{cen})$ of the driving source. The time-dependent linear differential equations are solved numerically through Runge-Kutta method, and amplitude of both plasmon modes are plotted as a function of excitation wavelength. Fig.\ref{Fig3} (a,b) shows plasmon mode amplitude driven by a plane wave for two different spectral bandwidths of CW light source. In both cases, we investigate the role of coupling strength between SPP and LSP modes and its effect on the spectral positions of hybrid plasmon modes. In Fig.\ref{Fig3} (a), the resonance wavelengths of SPP and LSP modes without coupling ($f_{p,l}=0$) are shown in red and blue curves at 702 nm and 600 nm respectively.
\begin{figure}[hbt!]
\includegraphics[scale=0.26]{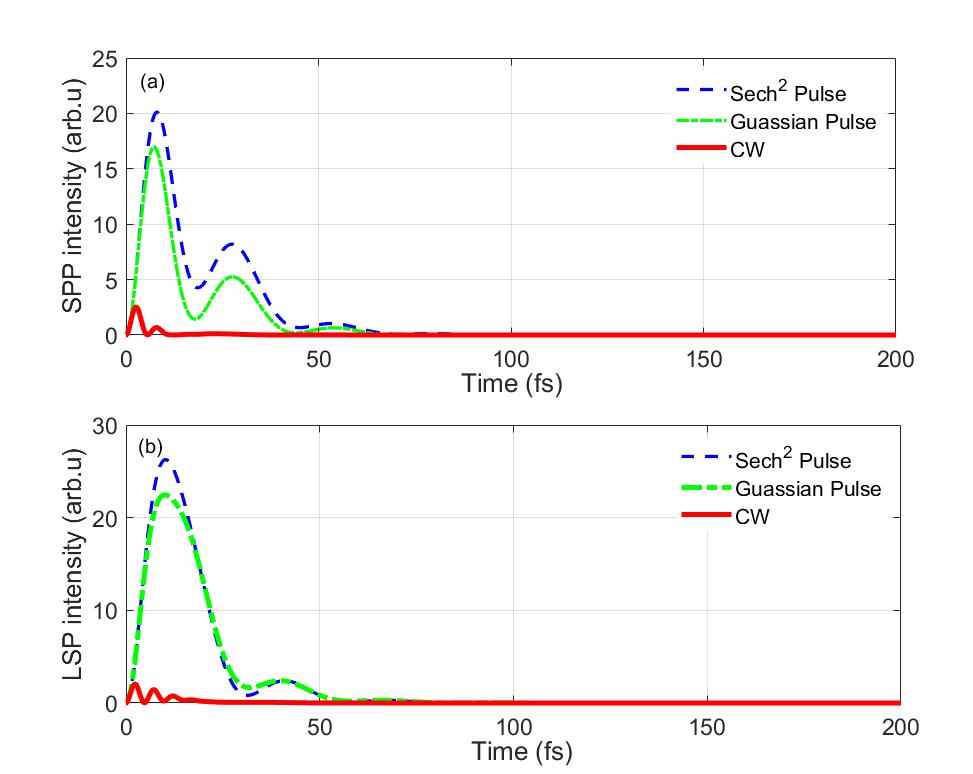}
\caption{\label{Fig4} SPP and LSP field intensities as a function of time for Gaussian, $Sech^2$ pulse and continuous wave (CW) excitation source with spectral parameters: $\omega_{p}= 1.07\omega_o$, $\omega_{l}= 1.25\omega_o$, $\gamma_{p,l}=0.04\omega_o$, $\tau=10 fs$ and $f_{p,l}=0.1\omega_o$.}
\end{figure}
Upon strong coupling, $f_{p,l}= 0.1\omega_o$, the resonance peaks shifts to the opposite directions with a significant broadening due to hybridization of different higher order modes. The two broad peaks of hybrid plasmon modes appear at 733 nm and 580 nm corresponding to SPP and LSP new peak spectral positions respectively. In the case of 750 nm excitation (Fig.\ref{Fig3} (b)), the hybridization of plasmon fields yields a blue shift in LSP resonance at 586 nm and red shift in SPP resonance at 707 nm. Moreover, the peak amplitude of SPP is more enhanced at 707 nm position as compared to $\lambda_{LSP}=$ 586 nm due to interference of in-phase SPP modes excited by $(\lambda_{cen})=$ 750 nm light.       
\section*{Effect of Pulse Shape and Temporal Width on Plasmon Resonances}
Here, we illustrate the effect of temporal shape and width of driving field on the plasmon mode intensity and oscillation time. We compute the response of the system from Eq.\ref{eq:4} and Eq.\ref{eq:5} and plot SPP and LSP field intensities as a function of time as shown in Fig.\ref{Fig4} (a,b) respectively.
In the case of CW excitation, the field intensity of both modes (SPP/LSP) is quite small which decays down shortly after 10 fs. In contrast to this, a 10-fold enhancement in the SPP mode intensity is observed for $Sech^2$ pulse excitation in Fig.\ref{Fig4} (a), and it is 8-fold for Gaussian pulse followed by a slowly damped oscillations until 60 fs. On the other hand, local dipole modes due to accumulation of energy at the rim of hole yield a maximum 13-fold enhancement in the field intensity for $Sech^2$ pulse case with a broad linewidth as shown in Fig.\ref{Fig4} (b). The increase in the mode intensities and oscillation period is mainly due to the high peak power of ultrashort pulse which stimulates highly energetic phase-locked plasmon polaritons at metal interfaces and around the hole edge simultaneously. Also, the carrier frequency of the envelope drives multiple SPP-Bloch modes with coherent polaritonic states and interference of these normal modes dramatically amplifies the plasmon intensity. 
\begin{figure}[hbt!]
\includegraphics[scale=0.42]{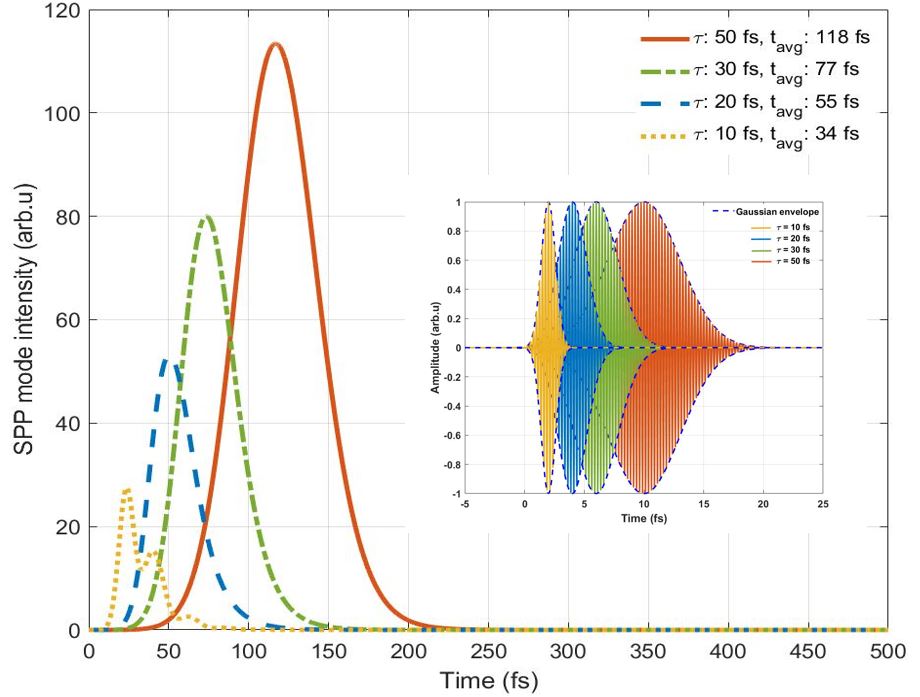}
\caption{\label{Fig5} Field-time profile of SPP for various pulse durations, inset shows the temporal waveforms of Gaussian pulses with different FWHM. The average lifetime of SPP oscillations increases linearly with pulse width. The parameters are taken as $\omega_{p}= 1.05\omega_o$, $\gamma_{p}=0.04\omega_o$ and $f_{p}=0.1\omega_o$}.
\end{figure} 
It is found that along with the enhancement in plasmon intensity, the linewidth of curve is also broadened in the pulse plasmon response. We think that this broadening results from the strong coupling of coherent and incoherent SPP and LSP modes within the nanoscale distances. Consequently, these oscillating modes undergo hybridization owing to different spectral phases. The resonant energy of each mode is affected by mixed hybrid states which broadens the linewidth of the intensity curve. Moreover, pulse duration which defines the envelope bandwidth leads to the temporal broadening of the plasmonic field.\\
To investigate the transient effect of pulse duration on the oscillation time of plasmon modes, we tune the FWHM of Gaussian envelope to various pulse durations. Fig.\ref{Fig5} shows the SPP field distribution as a function of time and evaluated average lifetime of SPP for respective pulse width $\tau$. The inset in Fig.\ref{Fig5} shows the temporal waveforms of Gaussian pulses with different pulse durations.
We evaluate the average lifetime $(t_{avg})$ of SPP/LSP modes by computing the eigen energies \cite{Yildiz2020lifetime, Asif2022} with total dissipated power \cite{Kirakosyan2016} as given in Eq.\ref{eq:6} where I(t) is the electric field intensity of plasmon mode and t is time.  
\begin{equation}
t_{avg}=\frac{\int_{0}^{\infty} t*I(t) dt}{\int_{0}^{\infty} I(t) dt}.
\label{eq:6}
\end{equation}      
For the plasmonic system, excited via CW source, the average lifetime of SPP resonances (red curve in Fig.\ref{Fig4} (a)) calculated from Eq.\ref{eq:6} is 10 fs. Whereas the SPP under the influence of 10 fs Gaussian pulse (Fig.\ref{Fig4} (a)), damps slowly with a lifetime of 34 fs. Further increase in the pulse duration changes the decay dynamics of plasmon oscillations drastically. The power spectra in Fig.\ref{Fig5} show the enhancement in the intensity and bandwidth of SPP modes for increasing pulse durations. For 20 fs pulse, the average oscillation time of propagating modes is 55 fs which exceed to 118 fs for 50 fs pulse. It is evident that pulse induced variations in the local density of polaritonic states \cite{Shahbazyan2016} dramatically changes the plasmon damping and drive the oscillations for longer period of time. With enhanced spatial and temporal characteristics features these SPPs modulate the properties of extraordinary transmission of light. To visualize the increase in the energy density and transmission intensity via coherent optical tuning, we perform 3D numerical simulations through finite difference time-domain (FDTD) method. 
\section{Computational model }
In the numerical simulations, we calculate the electric field flux in the near and far-field regime by using FDTD method which provides the time-dependent solutions of 3D Maxwell's equations and evaluates the power density spectra through the Poynting vector. The frequency and the time-domain spectra are obtained via standard Fourier transform of time-dependent field passing through the plane of EOT structure. We simulate the EOT structure, in the same way as modeled in Fig.\ref{Fig1}, a 100 nm thick Au film, with a circular hole of diameter 200 nm, layered on a glass substrate with a thin (2 nm) adhesion layer of Cr. The refractive indexes of dielectric background are taken as $n_g= 1.5$ for glass and $n_a= 1$ for air at the bottom and top surface of Au film, respectively. The parameters of frequency-dependent dielectric permittivity $\epsilon_m$ for Au, is taken from \cite{Rakic:98,Johnson1972}. For an infinite metal-dielectric surface perfectly matched layer (PML) is used in the z-direction, and periodic boundary conditions are used in the x and y directions. The FDTD simulation region, surrounds a unit cell of a two-dimensional hole array in the xy plane, defines the lattice constant $p$ of the array, which is taken as $400$ nm.\\
\begin{figure*}[hbt!]
\includegraphics[scale=0.58]{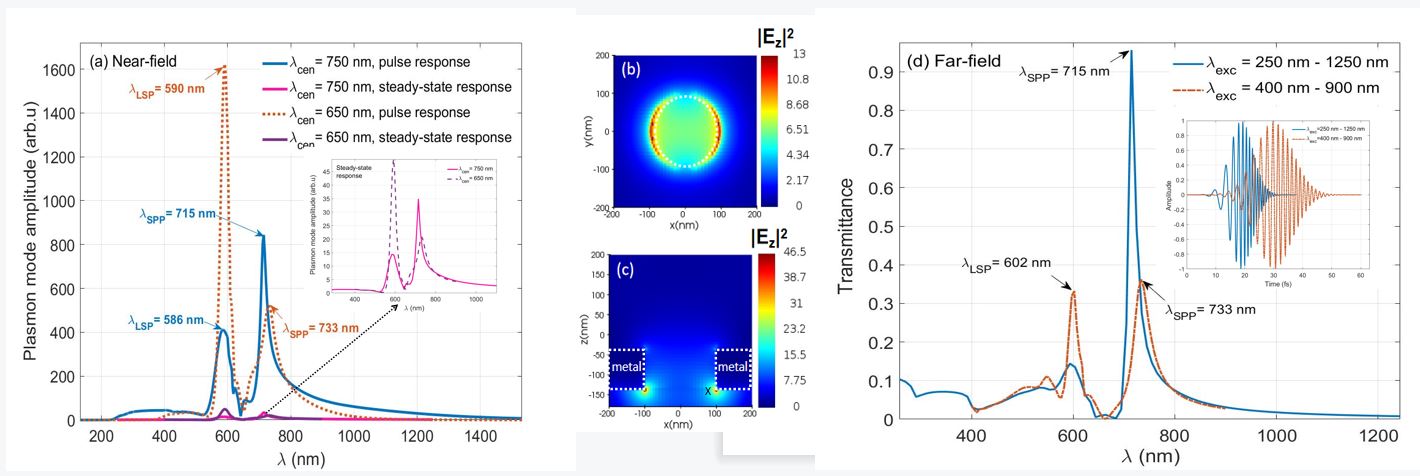}
\caption{\label{Fig6} (a) Electric field amplitude in the frequency domain near the hole edge (point X) for 750 nm and 650 nm excited pulses. The inset shows the field amplitude in the steady-state response of the system for the CW excitation. (b) Power density distribution on the xz and (c) xy-plane of the unit cell, for 750 nm pulse and the data is recorded at the peak transmitted wavelength 715 nm. The dashed lines indicate the metal boundaries. (d) Extraordinary transmission spectra of Au hole array system for two different spectral bands. Inset shows the temporal waveforms of Gaussian pulses.}
\end{figure*}
A Gaussian pulse, linearly-polarized in the x-direction and propagating in the z-direction, illuminates the EOT structure from the glass side, and the transmitted signal is collected from the metal-air side by placing a transmission monitor 200 nm above the metal surface. We compare the response of the system in the near and far-field regions for two different carrier wavelengths ($\lambda_{cen}$) of Gaussian pulse. We choose ($\lambda_{cen}$) such that; in the first case, pulse with a broad spectral band $(250 nm - 1250 nm)$ has $\lambda_{cen}: 750 nm$, duration $6 fs$ and pulse offset $18 fs$. In the second case, pulse with comparatively narrow bandwidth $(400 nm - 900 nm)$ has $\lambda_{cen}:$ 650 nm, $\tau :$ 10 fs and 30 fs offset. The pulse offset is taken larger than pulse duration to avoid pulse clipping in time. The simulations are performed by choosing the conformal mesh settings, and the simulation time is taken as 500 fs to ensure convergence. To evaluate the power flowing into the system, we place power monitors within the simulation region to investigate the transmission and electric field profiles in the frequency- and the time domain respectively. The frequency-domain power monitors are situated along the y and z planes to record the near-field distribution of LSP and SPP modes around the corner and above/below the metal-dielectric interface respectively.
\section*{Near and Far-Field Power Spectra of Plasmon Modes in the Frequency-Domain}
Here, we first calculate the plasmon mode amplitude near the hole edge (point X in Fig.\ref{Fig6} (c)) at the metal-glass interface for two different carrier wavelengths of the driving source. The spectra demonstrate the comparison of mode amplitudes in the pulse and steady-state response of the system as shown in Fig.\ref{Fig6} (a). The inset shows the zoom in view of the field amplitude in the steady-state calculated from the CW normalization, which provides the impulse response of the system normalized to the time-dependent driving pulse. The spectral positions and steady-state field amplitude of hybrid plasmon modes, for both driving sources, is in close agreement to our analytical results (Fig.\ref{Fig3}). In contrast to the steady-state, the pulse response shows a significant enhancement in the plasmon field intensity as predicted. Furthermore, we observe the change in the spectral positions of plasmon modes by tuning the carrier frequency of the driving field. For 650 nm pulse, the near-field spectrum shows two dominant peaks of LSP and SPP modes in the coupled state; the peak below the Rayleigh wavelength ($\lambda_{R}=$) 600 nm \cite{Rodrigo2016} i.e., 590 nm is associated to the localized dipole resonances ($\lambda_{LSP}$) whereas the peak appear at 733 nm results from the propagating surface modes ($\lambda_{SPP}$). Interestingly, due to strong coherent coupling of dipole modes with incoming photon energy, LSP field is more intense compared to SPP. However, for 750 nm pulse, the resonant energies (both modes) blue shift along with the change in peak amplitudes. In this case, propagating modes gain more strength due to the interference of phase-locked SPP modes which amplifies field amplitude at 715 nm in comparison to LSP (583 nm).\\
To elucidate the impact of spatial dynamics of LSP and SPP modes on the transmission properties, we analyze the near-field distribution, measured through 2D frequency domain monitors for 750 nm pulse, in the xy and xz-plane of the EOT structure. The power spectrum (Fig.\ref{Fig6} (b)) show an upside view of intense localized field around the hole at the metal-air interface. Fig.\ref{Fig6} (c) demonstrates that at the edges of cavity (metal-glass interface), the field amplitude gains strength due to the coupling of SPP and LSP modes in contrast to other parts of the metal plane. \\
We measure the extraordinary transmission spectra for two different broadband pulse excitations as shown in Fig.\ref{Fig6} (d), the inset shows the temporal profiles of Gaussian pulses. For wide-band (750 nm) pulse, the spectra indicate two EOT modes associated with $\lambda_{SPP}$ and $\lambda_{LSP}$ resonant wavelengths. At 715 nm, a three-fold enhancement in the peak transmittance occurs in comparison to the peak at 733 nm. This is due to large spectral bandwidth of Gaussian pulse which induces coherent excitation of multiple SPP-Bloch modes and allows superposition of phase-locked local (at the edge) and short-range (SR) propagating modes at the entrance and exit of the nanohole. Since the SR leaky modes present a large density of electromagnetic states therefore they easily couple to the broadband TM components of incident light. Furthermore, the cutoff wavelength is larger than the skin depth of metal, which makes the coherent SPP resonances strongly couple at both interfaces of the metal film. Both leaky and bound resonant states with coherent phases enhance the peak intensity of the EOT signal to 0.95 which corresponds to $95\%$ of transmission of incident light from the EOT structure. In contrast to this, the narrow band (650 nm) pulse yields two EOT modes with peak transmittance $0.35$ redshifts at 733 nm. The peak appeared close to $\lambda_{R}:$ 600 nm resonance due to the transmitted mode supported by the localized normal modes coupled to the free-space field at the first-order diffraction. The hybridization of both modes near the hole edge attributed to Fano resonance of non-radiative and radiative modes scattered from the hole and at the edge of circular cavity respectively \cite{Genet2003,luk2010fano}. 
\section*{Time-Resolved Spectra of Near-Field Plasmon Resonances}
\begin{figure}[hbt!]
\includegraphics[scale=0.22]{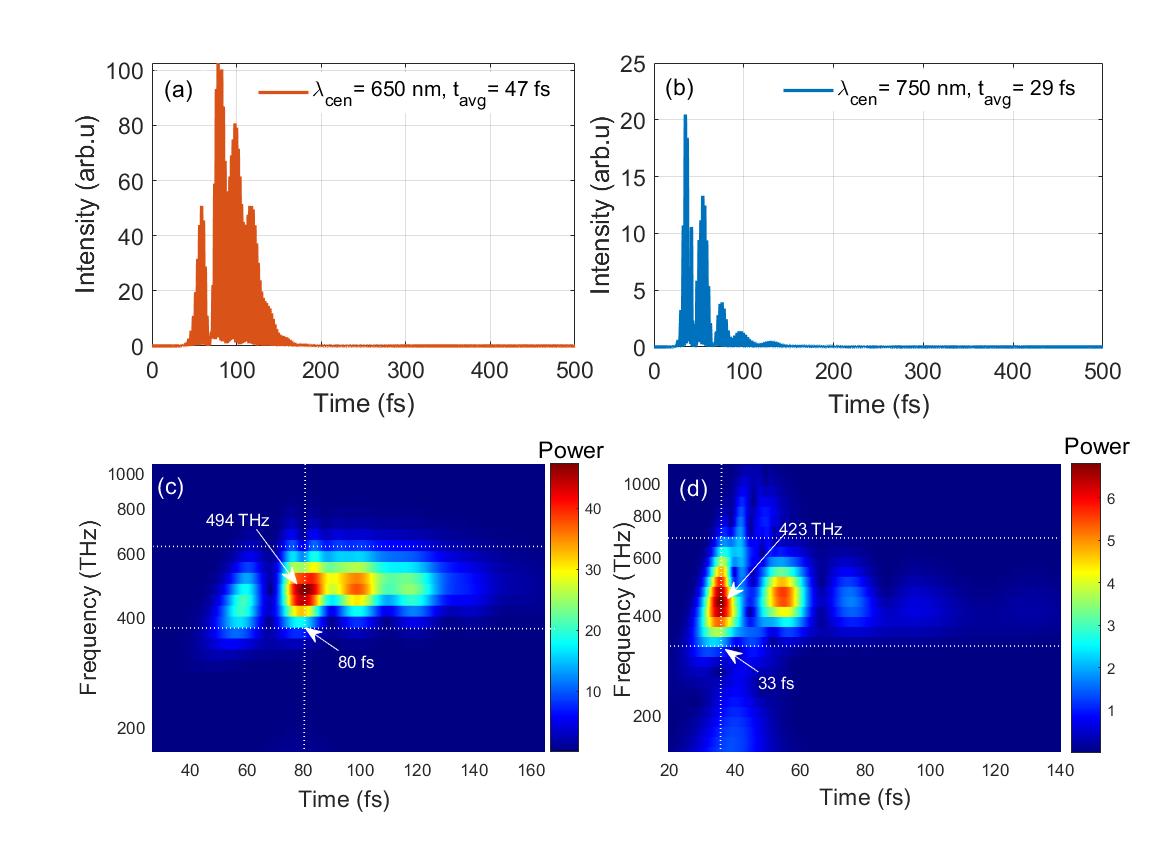}
\caption{\label{fig7} 
Electric field intensity as a function of time near the hole edge (point X) for (a) 650 nm pulse of $10$ fs duration and 30 fs offset time and (b) 750 nm broadband pulse with $\tau = 6 $ fs and 18 fs pulse offset. (c,d) Continuous wavelet transform (CWT) spectra for respective field-time profiles of plasmon modes.}
\end{figure}

In this section, we analyze temporal and spectral dynamics of plasmon modes based on field-time distribution and complex Morlet wavelet analysis. Generally, the spectroscopic features of plasmon modes in FDTD are derived through the frequency-domain spectrum or by taking the Fourier transform of the time-dependent electric-field signal. In the latter case, the power spectrum cannot uncover the time-domain character of each frequency component and its evolution with time. To understand the time-resolved spectral dynamics of plasmon mode, we use the continuous wavelet transform (CWT), which decomposes the complex spectral and temporal features of plasmonic signal in the time-frequency space within the timescale of coherent plasmon oscillations. This allows us to determine the dominant normal modes in the system and how those modes vary in time and contribute to the EOT signal.\\
A time-dependent plasmonic signal consists of a combination of modes oscillating with different frequencies. To extract local frequency information, we perform CWT on the electric field power spectrum obtained from the field-time monitor, placed near the hole edge at the metal-glass interface. Fig.\ref{fig7} (a,b) show the numerically simulated near-field power spectra of plasmon modes for 650 nm and 750 nm pulses. Each spectrum exhibits damped periodic oscillations of coupled plasmons following the impulsive excitation.
In Fig.\ref{fig7} (a), the plasmon oscillations with the maximum intensity around 80 fs correspond to the resonant LSP modes, excited at the peak envelope position which decays down shorty after 120 fs. Whereas, in Fig.\ref{fig7} (b), the wide spectral band of ultrashort pulse induces constructive interference between the accumulative pathways of SPP modes, which couples the local fields to outward propagating modes and results in the lower field strength of LSP near the hole edge. This shows a clear correlation with the peak amplitude at $\lambda_{LSP}$ in Fig.\ref{Fig6} (a), which is more intense in contrast to the LSP oscillations in 750 nm pulse. Now to determine the active resonant modes in the power spectrum, we perform CWT on the field-time profiles of Fig.\ref{fig7} (a,b) individually. For a time-domain signal E(t), the CWT $S(w, a)$, using the Morley wavelet method \cite{Deng2005} is defined as, 
\begin{equation}
S(a,w) = \frac{1}{\sqrt(a)}\int_{-\infty}^{\infty} E_z(t)\psi^{\ast}(\frac{t-w}{a})\,dx,
\label{eq:7}
\end{equation} 
where a is the scale factor, and $w$ is the time-dependent translation factor of wavelet transform (Eq.\ref{eq:7}), which depends on the frequency and the time of the time-frequency distribution, respectively. $\psi^{\ast}(\frac{t-w}{a})$ is the complex conjugate of Morlet wavelet function, which is defined as the product of complex electric field amplitude and Gaussian window given by,
\begin{equation}
\psi(t) = \frac{1}{(\pi)^{1/4}}e^{-i\omega_c t-\frac{t^2}{2\tau_{b}^2}},
\label{eq:8}
\end{equation} 
where $\tau_b$ is the bandwidth parameter, and $\omega_c$ is the central frequency of the wavelet. By choosing the scale and frequency parameters and using the CWT algorithm in the Matlab software, the time and frequency resolved CWT spectra are plotted, as shown in Fig.\ref{fig7} (c,d). The CWT spectrogram demonstrate the probability distribution of the instantaneous frequency of dominant resonant modes evolving in the system at an instant of time along with maximum power flowing into the system.
In Fig.\ref{fig7} (c), the dominant frequency band in the range from 430 THz to 529 THz, starts oscillating from 50 fs and decays down around 120 fs. In this band, 494 THz is the higher order mode corresponding to 608 nm wavelength with highest peak power appear at 80 fs which contribute to $35\%$ of the total field passing through the hole. In contrast to this, in Fig.\ref{fig7} (d), the mode distribution map shows a broad band of active dominant modes from 344 THz to 642 THz starting from 25 fs and ending up around 70 fs. In this case, the exchange of energies between the dominant resonant modes and the distant frequency bands amplify the coherent oscillation of propagating modes which give rise to intense EOT signal.
\begin{figure}[hbt!]
\includegraphics[scale=0.45]{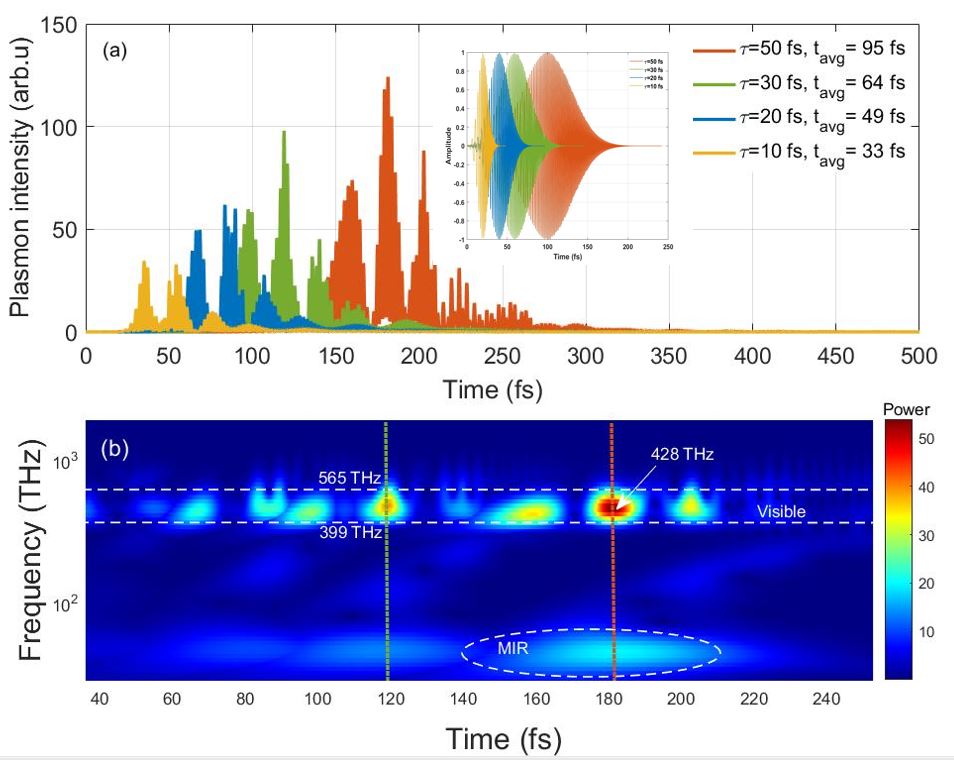}
\caption{\label{fig8} 
(a) Electric field power spectra as a function of time for different pulse durations measured at point X. Inset: Gaussian pulses for different $\tau$. (b) CWT spectrogram representing time-dependent distribution of instantaneous frequencies of resonant modes corresponding to each curve. }
\end{figure}
The higher order $TM_{01}$ mode, with 423 THz frequency and carrier wavelength of 709 nm, possesses the maximum intensity at 33 fs followed by a much faster decrease in the field amplitude. We investigate the fast decay dynamics of oscillating modes for both cases by calculating the average lifetime of plasmon modes from Eq.\ref{eq:6}. In Fig.\ref{fig7} (a), the average lifetime of plasmon modes excited by 10 fs Gaussian pulse is 47 fs. On the other hand, the power spectrum obtained for 6 fs pulse (Fig.\ref{fig7} (b)) indicates a rapid decay of plasmon modes, with an average lifetime of 29 fs. For the pulse duration shorter than the decay time of SPP, the excitation at the metal-glass interface is impulsive and the propagation dynamics changes according to oscillation strength of plasmon resonances. The difference in the oscillation time of plasmon modes in both cases stems from the temporal and spectral width of driving pulse. For instance, in 750 nm pulse, though the large spectral width excite multiple SPP modes in the system but due to small pulse duration no remarkable change is observed in the decay dynamics of plasmon modes.\\
We anticipate the change in the intensity and oscillation time of hybrid plasmon modes by varying pulse duration. Fig.\ref{fig8} (a,b) show the power spectra obtained for different values of $\tau$ (750 nm) measured near the hole corner and the corresponding CWT spectrogram representing the instantaneous distribution of dominant resonant modes. The inset shows the temporal profiles of driving pulses for different values of $\tau$. In Fig.\ref{fig8} (a), by increasing the pulse duration the field intensity not only increased gradually and but also the fast damping of plasmon modes decreased with enhanced periodic oscillations. For a 10 fs pulse, the average lifetime of plasmon mode oscillating near the hole edge increased up to 33 fs. 
\begin{figure}[hbt!]
\includegraphics[scale=0.35]{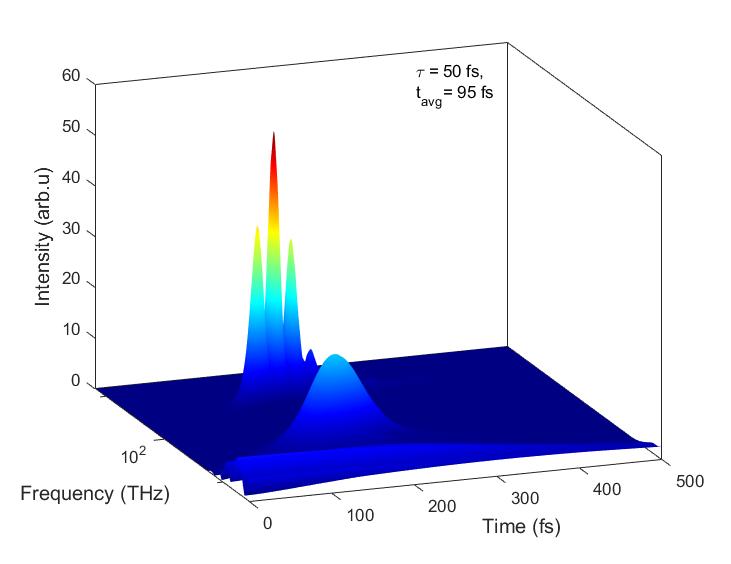}
\caption{\label{fig9} 
3D map of spectral distribution of plasmon modes in the time domain for 750 nm pulse and duration 50 fs. The average oscillating time of coupled resonant modes is 95 fs.}
\end{figure}
And as the pulse width gets broader, the decaying plasmons gain energy from the driving pulse, which derives it for longer time, depending on the duration of the pulse envelope. For 20, 30, and 50 fs pulses, the temporal modulation yields a significant enhancement in the average lifetime of plasmon modes in agreement with our analytical results obtained in Fig.\ref{Fig5}.\\
In Fig.\ref{fig8} (b), the CWT spectrogram demonstrates the dominant resonant modes in each spectrum of Fig.\ref{fig8} (a) and their evolution in time. The power spectra show that the spatial and temporal coherence of plasmon modes strongly depends on the temporal profile of the driving field. By varying pulse width, the increase in the pulse energy modulates the oscillation strength and decay length of plasmon modes. These modes with high field intensity and enhanced temporal resolutions are more confined locally in the visible spectral region (399 THz to 565 THz). In the case of a 50 fs pulse, the coherent interference of dominant resonant modes (428 THz to 459 THz frequencies), which occurs at 180 fs, contribute to the EOT signal. In this temporal window, a small band of long-range modes (with long decay lengths) also appeared in the mid-infrared (MIR) regime. The 3D illustration of periodic oscillations excited by 50 fs Gaussian pulse is given by a 3D scalogram as shown in Fig.\ref{fig9}, which demonstrates the time-dependent spectral distribution of local electric field components near the hole edge and their oscillating trends governed by higher and lower order SPP and LSP modes in the plasmonic system. 
\section{Conclusion}
We demonstrated all-optical control of ultrafast plasmon resonances, in the metal-nanohole structure, at the femtosecond time scale. We investigated the spectral characteristics of coupled plasmon modes (SPP and LSP) and analyzed the effect of ultrashort light pulse on the spectral and temporal characteristics of hybrid plasmon modes in contrast to conventional CW light source. The degrees of freedom for such optical control are temporal shape, duration, and spectral bandwidth of the ultrashort pulse. Our 3D FDTD results showed that the broad spectral bandwidth of the Gaussian pulse leads to a significant enhancement in the amplitude of strongly coupled SPP and LSP modes near the hole corner at the metal-glass interface. Also, a wide band (250 nm - 1250 nm) pulse excites multiple resonant SPP modes at both metal interfaces simultaneously, and the constructive interference of these in-phase modes amplifies local plasmon fields, which permit 95\% of the incident light at wavelength 715 nm to transmit through the subwavelength hole. We found that the peak power of Gaussian pulse not only increases the plasmon field intensity but also pulse duration enhances the oscillation time of the plasmonic field up to 100 femtoseconds. Moreover, modulating the pulse bandwidth allows the spectral tuning of EOT signal. The increase in the spatial and temporal response of SPP and LSP modes also incorporates in the lifetime enhancement of extraordinary optical transmission. Our approach provides a way to achieve all-optical control of the EOT device, which manifests its use as a biochemical sensor. It also enables on-chip active tuning of optoelectronic elements for enhanced spectroscopic and sensing applications. 
\section{Acknowledgement} 
R.S. and M.E.T acknowledge support from TUBITAK Project No. 121F030
\bibliographystyle{spphys}

\begin{thebibliography}{10}
\providecommand{\url}[1]{{#1}}
\providecommand{\urlprefix}{URL }
\expandafter\ifx\csname urlstyle\endcsname\relax
  \providecommand{\doi}[1]{DOI \discretionary{}{}{}#1}\else
  \providecommand{\doi}{DOI \discretionary{}{}{}\begingroup
  \urlstyle{rm}\Url}\fi

\bibitem{Dhama2022}
R.~Dhama, A.~Panahpour, T.~Pihlava, D.~Ghindani, H.~Caglayan, Nature
  Communications \textbf{13} (2022)

\bibitem{Taghinejad2019}
M.~Taghinejad, W.~Cai, ACS Photonics \textbf{6}, 1082 (2019)

\bibitem{Lechago2019}
S.~Lechago, C.~García-Meca, A.~Griol, M.~Kovylina, L.~Bellieres, J.~Martí,
  ACS Photonics \textbf{6}, 1094 (2019)

\bibitem{Zhang2015}
J.~Zhang, M.~Irannejad, M.~Yavuz, B.~Cui, Nanoscale Research Letters
  \textbf{10}, 1 (2015)

\bibitem{Liu2004}
Y.~Liu, J.~Bishop, L.~Williams, S.~Blair, J.~Herron, Nanotechnology
  \textbf{15}, 1368 (2004)

\bibitem{Lertvachirapaiboon2018}
C.~Lertvachirapaiboon, A.~Baba, S.~Ekgasit, K.~Shinbo, K.~Kato, F.~Kaneko,
  Biosensors and Bioelectronics \textbf{99}, 399 (2018)

\bibitem{Garcia-Vidal2010}
F.J. Garcia-Vidal, L.~Martin-Moreno, T.W. Ebbesen, L.~Kuipers, Reviews of
  Modern Physics \textbf{82}, 729 (2010)

\bibitem{Ebbesen1998}
T.W. Ebbesen, H.J. Lezec, H.F. Ghaemi, T.~Thio, P.A. Wolff, Nature 1998
  391:6668 \textbf{391}, 667 (1998)

\bibitem{Degiron2005}
A.~Degiron, T.W. Ebbesen, Journal of Optics A: Pure and Applied Optics
  \textbf{7}, S90 (2005)

\bibitem{Lee2015}
S.~Lee, H.~Song, S.~Hwang, J.~ryul Choi, K.~Kim, Handbook of Photonics for
  Biomedical Engineering pp. 1--22 (2015)

\bibitem{Chen2015}
S.~Chen, L.~Meng, J.~Hu, Z.~Yang, Plasmonics \textbf{10}, 71 (2015)

\bibitem{Sangiao2016}
S.~Sangiao, F.~Freire, F.D. León-Pérez, S.G. Rodrigo, J.M.D. Teresa,
  Nanotechnology \textbf{27} (2016)

\bibitem{Gehan2011}
H.~Gehan, C.~Mangeney, J.~Aubard, G.~Lévi, A.~Hohenau, J.R. Krenn, E.~Lacaze,
  N.~Félidj, Journal of Physical Chemistry Letters \textbf{2}, 926 (2011)

\bibitem{Barnes2006}
W.L. Barnes, Journal of Optics A: Pure and Applied Optics \textbf{8}, S87
  (2006)

\bibitem{Barnes2003}
W.L. Barnes, A.~Dereux, T.W. Ebbesen, Nature 2003 424:6950 \textbf{424}, 824
  (2003)

\bibitem{Yildiz2020}
B.C. Yildiz, A.R. Rashed, H.~Caglayan, Journal of Optics \textbf{22}, 065001
  (2020)

\bibitem{Sahin2020}
R.~Sahin, Optics Communications \textbf{454}, 124431 (2020)

\bibitem{Stockman2004}
M.I. Stockman, D.J. Bergman, T.~Kobayashi, Physical Review B - Condensed Matter
  and Materials Physics \textbf{69} (2004)

\bibitem{Roland2018}
R.~Müller, J.~Bethge, Physical Review B \textbf{98}, 085428 (2018)

\bibitem{Stockman2002}
M.I. Stockman, S.V. Faleev, D.J. Bergman, Physical Review Letters \textbf{88},
  67402/1 (2002)

\bibitem{qelife2013}
M.E. Taşgin, Nanoscale \textbf{5}, 8616 (2013)

\bibitem{Giannini2011}
V.~Giannini, A.I. Fernández-Domínguez, S.C. Heck, S.A. Maier, Chemical
  Reviews \textbf{111}, 3888 (2011)

\bibitem{Melikyan2004}
A.~Melikyan, H.~Minassian, Applied Physics B: Lasers and Optics \textbf{78},
  453 (2004)

\bibitem{Roland2003}
R.~Müller, V.~Malyarchuk, C.~Lienau, Physical Review B \textbf{68}, 205415
  (2003)

\bibitem{MacDonald2009}
K.F. MacDonald, Z.L. Sámson, M.I. Stockman, N.I. Zheludev, Nature Photonics
  \textbf{3}, 55 (2009)

\bibitem{Bahar2022}
E.~Bahar, U.~Arieli, M.V. Stern, H.~Suchowski, Laser and Photonics Reviews
  \textbf{16} (2022)

\bibitem{Baumert2007}
T.~Baumert, J.~Helbing, G.~Gerber, \emph{Coherent Control With Femtosecond
  Laser Pulses} (John Wiley and Sons, Ltd, 2007), vol. 101, pp. 47 -- 82

\bibitem{guenay_controlling_2020}
M.~Guenay, Z.~Artvin, A.~Bek, M.~Taşgin, JOURNAL OF MODERN OPTICS
  \textbf{67}(1) (2020)

\bibitem{Nehl2008}
C.L. Nehl, J.H. Hafner, Journal of Materials Chemistry \textbf{18}, 2415 (2008)

\bibitem{Yildiz2020lifetime}
B.C. Yildiz, A.~Bek, M.E. Tasgin, Physical Review B \textbf{101} (2020)

\bibitem{Asif2022}
H.~Asif, R.~Sahin, Journal of Optics (United Kingdom) \textbf{24} (2022).
\newblock \doi{10.1088/2040-8986/ac58a3}

\bibitem{Kirakosyan2016}
A.S. Kirakosyan, M.I. Stockman, T.V. Shahbazyan, Physical Review B \textbf{94},
  155429 (2016)

\bibitem{Shahbazyan2016}
T.V. Shahbazyan, Physical Review Letters \textbf{117} (2016)

\bibitem{Rakic:98}
A.D. Raki\'{c}, A.B. Djuri\v{s}i\'{c}, J.M. Elazar, M.L. Majewski, Appl. Opt.
  \textbf{37}(22), 5271 (1998)

\bibitem{Johnson1972}
P.B. Johnson, R.W. Christy, Physical Review B \textbf{6}, 4370 (1972)

\bibitem{Rodrigo2016}
S.G. Rodrigo, F.D. León-Pérez, L.~Martín-Moreno, Proceedings of the IEEE
  \textbf{104}, 2288 (2016)

\bibitem{Genet2003}
C.~Genet, M.P.V. Exter, J.P. Woerdman, Optics Communications \textbf{225}, 331
  (2003)

\bibitem{luk2010fano}
B.~Luk'yanchuk, N.I. Zheludev, S.A. Maier, N.J. Halas, P.~Nordlander,
  H.~Giessen, C.T. Chong, Nature materials \textbf{9}(9), 707 (2010)

\bibitem{Deng2005}
Y.~Deng, Z.~Wu, L.~Chai, C.Y. Wang, Z.~Zhang, Optics Express, Vol. 13, Issue 6,
  pp. 2120-2126 \textbf{13}, 2120 (2005)

\end{thebibliography}

\end{document}